\documentclass[aps,pra,showpacs,twoside,twocolumn,10pt]{revtex4-2}
\usepackage[colorlinks=true, citecolor=blue, urlcolor=blue ]{hyperref}
\usepackage{epsfig,newlfont,amssymb,amsfonts,amsmath,bm,subfigure,palatino,mathtools,amsthm,braket,soul,enumitem,graphics,graphicx,times,physics, multirow, makecell,diagbox}

\usepackage[normalem]{ulem}
\usepackage{tabularx}
\usepackage[table,xcdraw]{xcolor}
\usepackage{cleveref}

\begin{document}
\title{ Probing non-equilibrium physics through the two-body Bell correlator }

\author{Abhishek Muhuri$^{1,2}$, Tanoy Kanti Konar$^{1}$, Leela Ganesh Chandra Lakkaraju$^{1,3,4}$, Aditi Sen(De)$^{1}$}
\affiliation{$^1$ Harish-Chandra Research Institute, A CI of Homi Bhabha National Institute, Chhatnag Road, Jhunsi, Allahabad - 211019, India}
\affiliation{$^2$ Center for Quantum Science and Technology, International Institute of Information Technology Hyderabad, Gachibowli, Hyderabad, Telangana 500032, India}
\affiliation{$^3$ Pitaevskii BEC Center and Department of Physics, University of Trento, Via Sommarive 14, I-38123 Trento, Italy }
\affiliation{$^4$ INFN-TIFPA, Trento Institute for Fundamental Physics and Applications, Trento, Italy}

\begin{abstract}

Identifying equilibrium criticalities and phases from the dynamics of a system, known as a dynamical quantum phase transition (DQPT), is a challenging task when relying solely on local observables. We exhibit that the experimentally accessible two-body Bell operator, originally designed to detect nonlocal correlations in quantum states, serves as an effective witness of DQPTs in a long-range (LR) $XY$ spin chain subjected to a magnetic field, where the interaction strength decays as a power law. Following a sudden quench of the system parameters, the Bell operator between nearest-neighbor spins exhibits a distinct drop at the critical boundaries. In this study, we consider two quenching protocols, namely sudden quenches of the magnetic field strength and the interaction fall-off rate. This pronounced behavior defines a threshold, distinguishing intra-phase from inter-phase quenches, remaining valid regardless of the strength of long-range interactions, anisotropy and system sizes. Comparative analyses further demonstrate that conventional classical and quantum correlators, including entanglement, fail to capture this transition during dynamics. 
   
\end{abstract}
\maketitle

\section{Introduction}
\label{sec:intro}

Long-range order ~\cite{lieb1961,Kurmann_physicaA_1982,Prosen_PRL_2010,Sandvik_PRL_2010} enables the distribution of correlations across distant parts of a system, making it highly advantageous for numerous quantum technologies \cite{Solfanelli2023,monika2023bettersensingvariablerangeinteractions}. This order may emerge either at quantum criticality (cf. \cite{Hauke_2010_dipolar_devil_staircase}) in short-range interacting systems or in long-range (LR) interacting quantum spin models~\cite{Sachdev2011Apr, sadhukhan_arxiv_2021}. These LR systems have recently become a central topic of research, particularly for realizing analog quantum simulators \cite{diessel_prr_2023, Defenu2024Jun, defenu_prl_2018, defenu_prb_2019, solfanelli_arxiv_2024, defenu_arxiv_2024}, and due to their natural implementation in diverse physical platforms such as Rydberg atom arrays \cite{rydberg_review_experiments}, dipolar systems \cite{dipolar_longrange}, polar molecules \cite{cold_gas_long_range_review}, trapped-ion setups \cite{trapped_ion_1, trapped_ion_2, trapped_ion_3}, and cold atoms in optical cavities \cite{cold_atom_cavity_long_range_1, cold_atom_cavity_long_range_2}.  Furthermore, these systems exhibit several distinctive phenomena, including violations of the Lieb–Robinson bound \cite{PhysRevB.93.125128}, deviations from the area law \cite{area_law_longrange}, and the emergence of novel equilibrium and dynamical phases of matter \cite{phases_longrange, uhrich_prb_2020}.

Over the years, the study of nonequilibrium physics has gained significant importance from both technological and foundational perspectives. A central question in this domain is the identification of physical quantities that can, during real-time dynamics, distinguish between cases where the initial and final Hamiltonians, after a sudden quench, belong to different equilibrium phases and those where they remain within the same phase \cite{heyl_2013, Heyl_2018}. This phenomenon is known as a dynamical quantum phase transition (DQPT) \cite{heyl_2013,Heyl_PRL_2014,Heyl_2018,Sirker_PRB_2014,Schmitt_PRB_2015}. It has been shown that quantities such as the Loschmidt echo and the rate function can successfully identify quantum critical points in the transverse Ising model, though they fail in certain parameter regimes of the $XY$ and $XYZ$ models~\cite{Sirker_PRB_2014, Vajna_PRB_2014}. Additionally, it was found that standard deviations of multipartite entanglement \cite{Stav_PRB_2020, Nicola_PRB_2022},  time-ordered correlation functions of finite length \cite{heyl_2018_OTOC}, and mutual information \cite{lakkaraju2023frameworkdynamicaltransitionslongrange,bhat_prb_2024} can effectively signal DQPT. DQPTs are also studied in the context of floquet driving \cite{jafri_floquet_dqpt_pra_22, jafri_floquet_dqpt_pra_2021_6, jafri_floquet_dqpt_prb_2022_1, jafri_floquet_dqpt_prb_2022_2, jafri_floquet_dqpt_xy_model_prb_2020_5}, across topological phases \cite{jafri_topo_dqpt_prb_2021_4, lakkaraju_dqpt_topo} and non-Hermitian exceptional points \cite{keshav_non_herm_DQPT}, long-range systems \cite{heyl_long_range_2018, jafri_longrange_dqpt_iop_2020_7,lakkaraju2023frameworkdynamicaltransitionslongrange}. Experimental realizations of DQPTs have also been achieved across a variety of platforms, including trapped ions~\cite{trapped_ion_dqpt}, nuclear magnetic resonance setups~\cite{nmr_dqpt}, superconducting circuits~\cite{guo_superconduting_dqpt}, and atoms in optical lattices~\cite{souvik_experiment,cold_atom_fermionic_dqpt}.

Conventional dynamical quantifiers used to identify DQPTs often demand complex measurement techniques, such as full state tomography. To overcome this limitation, we propose an experimentally accessible observable, the Bell operator \cite{Horodecki_PLA_1995}, which, beyond its foundational role, can efficiently capture DQPT signatures. Since its introduction \cite{Bell64, CHSH}, Bell nonlocality has motivated extensive research aimed at characterizing quantum correlations in many-body systems~\cite{Tura_science_2014, Scarani_science_2016, bruner_nonlocality_rev, Engelsen_PRL_2017, Tura_PRX_2017, Piga_PRL_2019, Wagner_PRL_2017, frrerot_prl_2021}, and quantifying the depth of nonclassical correlations, that is, the number of parties sharing genuinely quantum correlations~\cite{Baccari_PRA_2019, Aloy_PRL_2019}. 

Extending this perspective, we employ the two-body Bell Clauser–Horne–Shimony–Holt (CHSH) correlator \cite{Bell64, CHSH} as a practical tool to identify quantum criticality in dynamical regimes. A further departure from conventional approaches lies in our consideration of the extended long-range  $XY$ model, an integrable system that simultaneously incorporates anisotropy, long-range interactions, and an external magnetic field. In this framework, quenches are implemented through abrupt changes in either the magnetic field strength or the interaction fall-off rate, both of which jointly determine the system’s magnetic phase. Starting with the ground state of the LR model and subjecting it to a sudden quench, we observe that the steady-state value of the Bell operator exhibits a pronounced drop whenever the system traverses a phase boundary. This distinct behavior enables the identification of a threshold value above which the quench remains confined to the same phase. We note that there have been previous studies that incorporate quasilocal string operators \cite{bandyopadhya_prl_2021} to identify DQPTs and also experimentally realized \cite{souvik_experiment} in the nearest-neighbor Ising model. We emphasize our study also incorporates anisotropy, long-range and a two-body quantifier. 

For a quantitative assessment, we define two key measures, the critical threshold value of the Bell correlator and its corresponding efficiency. Our analysis reveals that, for both types of quenches, 
the critical values of the Bell correlator follow a Gaussian-like law. Moreover, the Bell correlator exhibits significantly higher efficiency than other classical correlators and bipartite entanglement measures, independent of the anisotropy parameter. This result highlights the robustness and versatility of the Bell operator as a local and experimentally accessible probe of many-body criticality in nonequilibrium quantum systems.

The paper is organized as follows. In Sec. \ref{sec:bellbehavior}, we introduce the LR Hamiltonian  and outline the computation of the Bell-CHSH operator for this model, along with its temporal behavior under field and coupling quenches. We further compare its steady-state and time-averaged values in this section. In Sec. \ref{sec:Belloperatorhquench}, we propose a novel DQPT quantifier based on the steady-state Bell operator and define a corresponding measure of efficiency. The section also discusses the functional dependence of this quantifier. Sec. \ref{sec:bell_vs_ent_and_correlator} presents a comparative study of the efficiencies of entanglement- and classical correlator-based quantifiers  with the proposed Bell-based approach. Sec. \ref{sec:exper} discusses the benefit of the proposed quantity from the experimental point of view over the existing ones in literature.  We finally conclude in Sec. \ref{sec:conclu}.

\section{Patterns of Bell correlator in dynamics across magnetic phases}
\label{sec:bellbehavior}

Before proposing the two-party Bell correlator as a reliable quantifier to capture equilibrium physics from the properties of a dynamical state, it is essential to first investigate and understand how this correlator behaves throughout the course of the system’s evolution. We introduce the system under study, outline the formalism required to analytically compute the Bell correlator, and examine its time-dependent behavior.

\subsection{Long-range extended $XY$ model and setup for dynamics}
\label{subsec:model}

The Hamiltonian \(N\) interacting spin-\(1/2\) systems that describes the long-range extended \(XY\) model  in the presence of a transverse magnetic field having the periodic boundary condition (PBC) (i.e., \(\sigma_{N+1}\equiv \sigma_1\)) reads as
\begin{eqnarray}
    \nonumber H&=&\sum_{j=1}^{N} \sum_{r=1}^{\frac{N}{2}} -\frac{J_r^\prime}{\mathcal{N}} \Bigg [\frac{1+\gamma}{4}\sigma_j^x\mathbb{Z}_r^z\sigma_{j+r}^x\nonumber+\frac{1-\gamma}{4}\sigma_j^y\mathbb{Z}_r^z\sigma_{j+r}^y\nonumber\\&&-\frac{h'}{2}\sum_{j=1}^{N}\sigma_j^z,
    \label{Eq.Hamiltonian}
\end{eqnarray}
where $\gamma$ is the anisotropy parameter,  \(h'\) is the strength of the external magnetic field, $\mathbb{Z}_r^z = \prod_{l=j+1}^{j+r-1}\sigma_l^z$, with $\mathbb{Z}_1^z=\mathbb{I}$
and \(\sigma^k\)(\(k=1,2,3\))s are the Pauli matrices, and \(J_r'=\frac{J}{r^\alpha}\) with \(\alpha\) being the fall-off rate, representing the strength of the power-law decay of the model and \(J>0\). Here \(\mathcal{N}=\sum_{r=1}^{N/2}\frac{1}{r^\alpha}\) is known as the Kac-scaling factor \cite{Kac_jmp_1963}, providing extensivity of the energy in the case of finite-size systems.  We set \(h=h'/J\) to make the analysis dimensionless. The critical points of the system are located at $h_c = -1+2^{1-\alpha}$ and $h_c^{(2)} = 1$.

In the free-fermionic picture, the Hamiltonian in terms of fermionic operators, \(c_j\) and \(c_j^{\dagger}\), takes the form 
\begin{eqnarray}
   \nonumber H&=&\sum_{n}\sum_{r}\frac{J_r}{2}\left (c_n^\dagger c_{n+r}+\gamma c_n^\dagger c_{n+r}^\dagger + \text{h.c}\right )\\&&+h(c_n^\dagger c_n-1/2).
   \label{eq:JW_hamil}
\end{eqnarray}
With the application of the Fourier transformation of the fermionic operators, \(c_j\) and \(c_j^{\dagger}\) as \(c_j = \frac{1}{\sqrt{N}}\sum_p e^{-i \phi_p j} c_p\), it gets further simplified as \(H = \sum_{p>0}^{N/2} H_p\) with \(p \in [-N/2, N/2]\) being the momentum (see Appendix \ref{App:JW_diagonalization}) with \(H_p\) being the \(4\times 4\) diagonalizable matrix having two blocks. Writing \(H_p\) in the momentum basis, the eigenspectra at finite temperature and the evolution of the system can be investigated. 

{\it Classical correlators of the dynamical states.} Initialization of the system to the ground state with the magnetic field strength \(h_i\) at \(t=0\),  the system is quenched by changing the magnetic field from $h_i$ to $h_f$ at \(t>0\). Hence, the evolved state of the \(N\)-party takes the form \(|\psi (t) \rangle = \exp(-i H(h_f) t) |\psi_0\rangle  \equiv U(t)  |\psi_0\rangle\), where \(|\psi_0\rangle\) is the ground state of the Hamiltonian and \(U(t)\) is the evolution operator. The bipartite reduced density matrix between two nearest-neighbor sites, \(i\) and \(i+1\), of the dynamical state can be obtained by tracing out all the parties except two parties, \(i\) and \(i+1\). Due to translational symmetry, all the reduced density matrices \(\rho_{ij}\) are the same for a fixed \(|i -j|\) and hence without loss of generality, we compute \(\rho_{12}(t)\) which is a function of magnetization, \(m^z_1(t)\) (\(m^z_2(t)\)) and classical correlators, \(C^{kl}(t) = \Tr(\tau^{kl} \rho_{12}(t))\) with \(k, l = x, y, z\), where $\tau^{kl}=\sigma^k \otimes  \sigma^l$.  The diagonalization procedure discussed above and in Appendix \ref{App:JW_diagonalization} enables us to write \(H(h_f)\) involved in the unitary operator in the momentum space. Therefore, the correlators reduce to 
\begin{eqnarray}
    C^{kl}(t) =\frac{2}{N}\sum_{p>0}\Tr[\tau^{kl}_p \left(U_p(h_f,t)\rho^p_0 U_p^{\dagger}(h_f,t)\right) ], 
\end{eqnarray}
where the unitary operator $U_p(h_f,t) = exp[iH_p(h_f)t]$, the initial two-party reduced state $\rho^p_0$, and the Pauli operators $\tau^{kl}_p$ are written in momentum space. 

Instead of altering the magnetic fields at \(t=0\) and \(t>0\), we also consider a situation where the initial Hamiltonian has the fall-off rate, \(\alpha_i\) while the evolution operator, \(U\) has the fall-off rate \(\alpha_f\). It is important to stress here that although the original DQPT-idea was based on the magnetic field quench where the critical points are based on the gap-closing, it may happen sometimes that the change of fall-off rates can also alter the system's energy gaps, due to which, the critical line $h_c = -1+2^{1-\alpha}$ depends on the the fall-off rate. Thus, it is possible to observe the gapless phase while quenching the coupling parameter $\alpha$, defining the magnetic boundary in the quenched parameter space. Hence, one may expect to capture this boundary through a suitable dynamical quantifier. In Sec. \ref{sec:Belloperatorhquench}, we will illustrate that this is indeed the case when one analyzes the Bell-CHSH operator in the dynamical setting.  

\subsection{Computation of Bell-CHSH operator for this model} 

Let us provide a summary of the Clauser-Horne-Shimony-Holt (CHSH) correlator, which is the simplest Bell scenario, involving two measurement settings for two parties with two outcomes, denoted by \((2,2,2)\).  In particular, consider that two observers, Alice and Bob,  sharing a bipartite state $\rho_{12}$,  can choose between two possible measurement settings, $\{A_1,A_2\}$ for Alice and $\{B_1,B_2\}$ for Bob, which are typically dichotomic, having two outcomes $\pm 1$. The Bell  CHSH operator \cite{Bell64, CHSH}  is defined as \begin{equation}
    \mathcal{B} = |\langle A_1 B_1\rangle + \langle A_1 B_2\rangle + \langle A_2 B_1\rangle - \langle A_2 B_2\rangle|,
    \label{Belloperator}
\end{equation}
where $\langle M\rangle = \Tr( M \rho_{12})$ is the expectation value of $M$ with respect to $\rho_{12}$. The correlator is upper bounded by  $2$ for local-hidden variable theories \cite{Bell64, CHSH} while the violation of this inequality detects the signature of nonlocal correlations. 

For a two-qubit state, \(\rho_{12}\), it can be written as \(\rho_{12} = \frac{1}{4}[I^4 + \sum_{k=x, y, z} m^k I^2\otimes \sigma^k + \tilde{m}^k \sigma^k \otimes I^2) + \sum_{k,l={x,y,z}}C^{kl} \sigma^k \otimes \sigma^l] \), where \(I^d\) denotes the identity operator in \(d\)-dimension.  The elements, \(C^{kl}\) form a \(3 \times 3 \) correlation matrix \(T\). It was shown that the Bell CHSH operator \cite{Horodecki_PLA_1995}, \(\mathcal{B}\), can be obtained by finding two largest  eigenvalues $\lambda_1$, and $\lambda_2$ of $\mathcal{M}:=T^T T$, i.e., \(\mathcal{B}  = 2 \sqrt{\lambda_1 + \lambda_2}\). 

In the case of the extended \(XY\) model, the two-party reduced density matrix of the evolved state possesses nonvanishing \(m^z, C^{ii}, C^{xy}\) and \(C^{yx}\)
with (\(i=x, y, z\)) and of the form of $X$-state. 
Hence, \(\lambda_1\) and \(\lambda_2\) involved in \(\mathcal{B}\) operator depend on five non-vanishing elements of \(T\)-matrix. Diagonalizing \(\mathcal{M}\), we find  three eigenvalues, \(\lambda_{\pm}\) and \((C^{zz})^2\), such that 
    \begin{align}
        2\lambda^2_{\pm} &= (C^{xx})^2 +(C^{yy})^2+(C^{xy})^2+(C^{yx})^2 \nonumber \\ 
        &\pm\sqrt{(\mathcal{C}_+^{aa}+\mathcal{C}_-^{ab})(\mathcal{C}_-^{aa}+\mathcal{C}_+^{ab}}),
    \end{align} 
where $\mathcal{C}^{aa}_\pm = (C^{xx}\pm C^{yy})^2 $ and $\mathcal{C}^{ab}_\pm = (C^{xy}\pm C^{xy})^2 $. 

Consequently, the Bell correlator for the time-evolved state becomes 
\begin{eqnarray}
    \mathcal{B}(\rho(t)) \equiv \mathcal{B} &=&  2 \sqrt{\max[\lambda_++\lambda_-,\lambda_++(C^{zz})^2]}. \nonumber\\
    \label{Eq.Belleigs}
\end{eqnarray}
Depending on the system parameters, the maximum value changes for the steady state, i.e., for a large time, after the initial fluctuations in the transient regime. 

\subsection{Dynamics of  Bell CHSH correlator}
\label{sec:dynamics}

The pattern of the \(\mathcal{B}\) operator exhibits some universal behavior, either when the initial and magnetic fields are chosen from the same or different phases. 

\begin{figure}
    \centering
    \includegraphics[width=\linewidth]{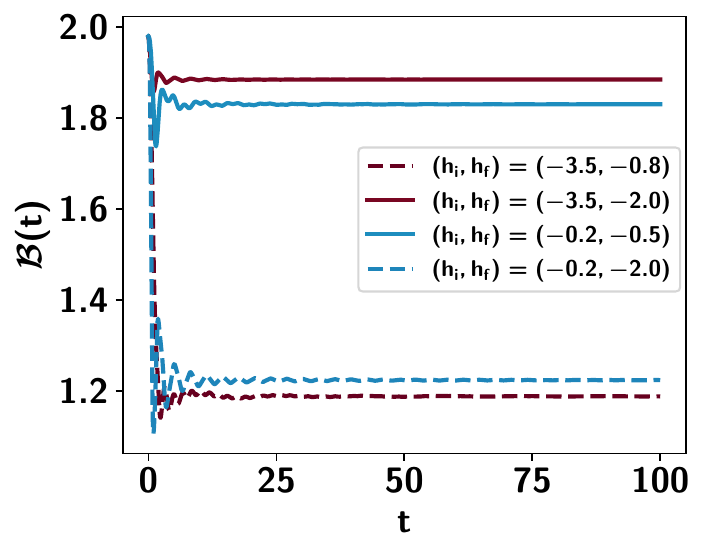}
    \caption{(Color online) {\bf Temporal behavior of Bell CHSH correlator under field quench. }
     The behavior of $\mathcal{B}(t)$ (ordinate) against time (abscissa) for different pairs of initial and quenched magnetic field strengths, (\(h_i, h_f\)), of the nearest-neighbor transverse Ising models (\(\gamma =1, \alpha =10\)). Solid lines (dark and light) represent when the initial and final fields  belong to the same phase while dashed lines (dark and light) are for inter-field quenches. We notice that intra-field quenches produce a higher saturated \(B(t)\) as compared to the inter-field quenches. Based on these observations, we introduce the physical quantity, \(\mathcal{B}_s(t)\) in Eq. (\ref{eq:Bsat}) to probe DQPT.    
    All axes are dimensionless.}
    \label{Fig:BelldropObs}
\end{figure}

{\it Effectiveness of Bell correlator with magnetic field quench.} Let us first investigate the case with the sudden quench of the magnetic field when the Hamiltonian involves nearest-neighbor interactions, i.e.,  with high \(\alpha\). 
\begin{enumerate}
    \item Starting from the ground state of the system with magnetic field, \(h_i\), the system is quenched to \(h_f\) resulting in the oscillation of \(\mathcal{B}\) with time before converging to a fixed value at large time (see Fig. \ref{Fig:BelldropObs} with \(\alpha =10\)).
    \item Importantly, when the initial and the final magnetic fields belong to the same magnetic phases, the steady-state value of \(\mathcal{B}(t)\) (as \(t \rightarrow \infty\)) is higher compared to the situation when initial and final magnetic fields belong to different phases, irrespective of the paramagnetic and ferromagnetic phases. In other words, we have the following observation:

 {\bf Observation 1.} {\it Under a sudden global quench, the Bell correlator of the initial state approaches a markedly lower steady-state value when the system is driven into a different phase, compared to quenches performed within the same phase.}

 \item  The initial values of \(\mathcal{B}\) and its dynamical values, both in the transient and steady state domains, cannot go beyond the value of \(2\), otherwise the monogamy of nonlocality will be violated~\cite{kuzy_prl_2011}. Hence, this analysis reveals that even though we are unable to ensure its nonlocal characteristics, we can have a measurable quantity that can identify the equilibrium physics of the system from its dynamics.
\end{enumerate}

\begin{figure}
    \centering
    \includegraphics[width=\linewidth]{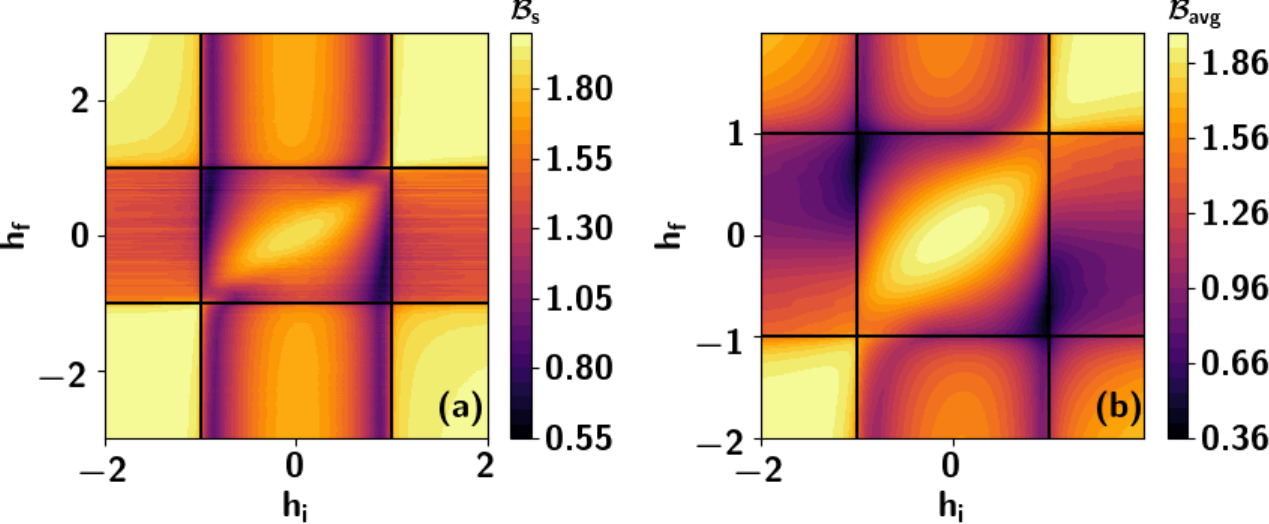}
    \caption{(Color online) {\bf Comparison between steady-state  and time-averaged Bell correlators across dynamical phases of the nearest-neighbor transverse Ising model. }
    (a) Map plot of the steady-state value, $\mathcal{B}_{s}$,  over the parameter space defined by $h_i$ (abscissa) and $h_f$ (ordinate) with (\(h_i, h_f\)).  Regions where $h_i$ and $h_f$ lie in different phases exhibit reduced $\mathcal{B}_{s}$, while quenches within the same phase yield elevated values.
    (b) Map plot for the time-averaged Bell correlator, $\mathcal{B}_{avg}$, exhibiting similar behavior to $\mathcal{B}_{s}$.
    All axes are dimensionless.}
    \label{Fig:Bell_satVSavg}
\end{figure}

Although it is known that equilibrium Bell correlation can efficiently detect the phase boundaries~\cite{justino_pra_2012,Piga_PRL_2019}, Bell correlation is known to be non-ergodic, i.e., its equilibrium and steady-state values do not coincide~\cite{Batle_PRA_2010}. It is interesting to check whether it can still carry the signature of the equilibrium phase transitions when the system is out of equilibrium.  The above analysis possibly indicates that it is indeed the case, emphasizing its ability to detect the quantum phase transition even in the dynamical setting. The two possible quantities that emerge from the behavior of \(\mathcal{B}\) with time is the saturated Bell operator and the average Bell operator, with averaging being performed over time, given respectively by
\begin{eqnarray}
    \label{eq:Bsat}
     \mathcal{B}_{s} &=& \lim_{t\rightarrow\infty} \mathcal{B}, \\
     \label{eq:Bavg}
\text{and} \,\,  \mathcal{B}_{avg} &=& \underset{t\to\infty}{\lim}\frac{1}{t}\int \mathcal{B} \, dt,
\end{eqnarray}
where \(\mathcal{B}\) operators are functions of the time and system parameters, \(h_i\), \(h_f\), \(\gamma\), and  \(\alpha\) and the integration is performed over a long time or until the time for which the system reaches steady state. Interestingly, we find that both quantities can capture the quantum critical points via dynamics (see Fig. \ref{Fig:Bell_satVSavg}). However, we notice that \(\mathcal{B}_s\) identifies the transition far more distinctly than \(\mathcal{B}_{avg}\), particularly in cases where the \((h_i, h_f)\)-pair lies within the same phase yet crosses a critical point during the quench.  Note that the previously known quantity for capturing DQPT requires a seven-body operator instead of the two-body operators considered here \cite{bandyopadhya_prl_2021, souvik_experiment}. Furthermore, it is important to stress that, due to the experimental breakthroughs, such operators are also accessible in laboratories without full-state tomography ~\cite{Christandl_PRL_2012,Rippe_PRA_2008,Stricker_PRXQuantum_2022,Cramer_NatureComm_2010}. 

It should be noted that, henceforth, the explicit time dependence $t$ is omitted from all indicators, including the Bell parameter, the classical correlators, and the entanglement measure, as the discussion hereafter pertains exclusively to their averaged or saturated values.

\section{Efficacy of dynamical Bell correlator in capturing equilibrium physics}
\label{sec:Belloperatorhquench}

A central question in assessing the usefulness of two-body correlators is whether they retain the signatures of the underlying equilibrium phases when the system is driven far from equilibrium. To address this, we investigate the long-time saturated value under sudden quenches of the transverse field. Two qualitatively distinct situations naturally emerge: (i) the pre- and post-quench parameters belong to the same equilibrium phase, and  (ii) the quench drives the system across a phase boundary. The results below demonstrate that this dichotomy is sharply imprinted in the dynamical behavior of the Bell operator.

In the present analysis, we focus on a two-parameter manifold, namely \((\alpha, h)\), and examine the dynamical response under a sudden quench from an initial configuration \((\alpha_i, h_i)\) to a final configuration \((\alpha_f, h_f)\). Two distinct quench protocols are considered for clarity: (i) a \emph{field quench}, where the coupling \(\alpha\) is held fixed while the transverse magnetic field is varied, i.e., \((\alpha, h_i) \rightarrow (\alpha, h_f)\); and (ii) a \emph{coupling quench}, wherein the magnetic field remains constant and only \(\alpha\) is altered, i.e., \((\alpha_i, h) \rightarrow (\alpha_f, h)\).  In what follows, we detail the procedure employed to perform phase detection and systematically assess the efficacy of the Bell correlator as a diagnostic observable, i.e., quantifying its sensitivity across distinct regions of the phase diagram.

\subsection{Performance quantifiers for quench across the parameter doublet $(\alpha, h)$}
\label{sec:Bellcorrpattern}

We partition the parameter space into two distinct dynamical regimes, corresponding to quenches performed within the same equilibrium phase and those traversing across different phases. The steady-state Bell correlator exhibits systematic variations across these regions, thereby providing a clear diagnostic of the phase structure. In particular, the regions where the quench remains confined to a single phase (effectively the off-diagonal regions of the quench parameter space in all figures, additionally the diagonal sections for specifically field quenches), are characterized by elevated values of the steady-state Bell correlator $\mathcal{B}_{s}$. In contrast, the remaining domains corresponding to quenches in different equilibrium phases yield markedly lower values.  

To encapsulate these distinctions more rigorously, we introduce the following useful quantity. \\
\textit{Critical benchmarking threshold (\(\mathcal{B}_c\)) --  When the steady-state Bell correlation $\mathcal{B}_s$ is higher than \(\mathcal{B}_c\), i.e.,  \(\mathcal{B}_s \geq \mathcal{B}_c\), it can unambiguously be inferred that the ensuing evolution remains confined within the same dynamical phase.}

As discussed previously, the structure of \(\mathcal{B}_c\) exhibits a nontrivial functional dependence on the underlying system parameters. Specifically, (i) under a \textit{field quench}, \(\mathcal{B}_c\) assumes a Gaussian profile whose width and amplitude are governed by the anisotropy parameter and the coupling coefficient; (ii) conversely, for a \textit{coupling quench}, \(\mathcal{B}_c\) displays a tri-Gaussian morphology that now depends simultaneously on both the field and anisotropy parameters. 

We define this efficiency obtained through a quantifier, \(\mathcal{Q}\) as the ratio between the area \(\mathcal{A}\) of the phase diagram distinguished by the benchmarking value \(\mathcal{Q}_c\) and the total area \(\mathcal{A}_{\mathrm{same}}\) corresponding to the same-phase quenches via that quantifier. Mathematically, this is expressed as
\begin{equation}
\eta_{\mathcal{Q}} = \frac{\mathcal{A}}{\mathcal{A}_{\mathrm{same}}}, ~\text{where} ~ 
\mathcal{A} = \left| \{ (q_i, q_f) \, | \, \mathcal{Q}_s \geq \mathcal{Q}_c \} \right| (\Delta q)^2.
\end{equation}
Here, \(q_i\) and \(q_f\) denote the initial and final values of the quenching parameter, respectively, while \(\Delta q\) represents the interval between consecutive values of \(q\). The symbol \(|\cdot|\) denotes the cardinality of the corresponding set. In this work, we use Bell correlator, bipartite entanglement and classical correlators as quantifiers to assess the efficiency. 

For a magnetic field quench, the field parameters \(h_i\) and \(h_f\) vary in the range \(-3 \leq h \leq 3\) with an interval \(\Delta h = 0.01\). Consequently, the total area associated with the same-phase quenches is given by
\begin{equation*}
\mathcal{A}^h_{\mathrm{same}}(\alpha) =  20 + 2^{3-\alpha} + 2^{3-2\alpha}.
\end{equation*}
Similarly, for a coupling quench, the coupling parameters \(\alpha_i\) and \(\alpha_f\) are varied from \(0.5\) to \(3.0\) with an interval $\Delta \alpha = 0.01$, yielding
\begin{equation*}
    \mathcal{A}^{\alpha}_{\mathrm{same}}(h) = 4.25 + 3\log_2(1+h) + 2[\log_2(1+h)]^2.
\end{equation*}
 Here, it is instructive to note that we use a similar set of efficacy parameters,  named $\eta_C$ and $\eta_\mathcal{E}$ corresponding to classical correlator $C^{zz}$ and entanglement of the steady state, respectively, to distinguish the phases dynamically in the next section. 

\begin{figure}
    \centering
    \includegraphics[width=\linewidth]{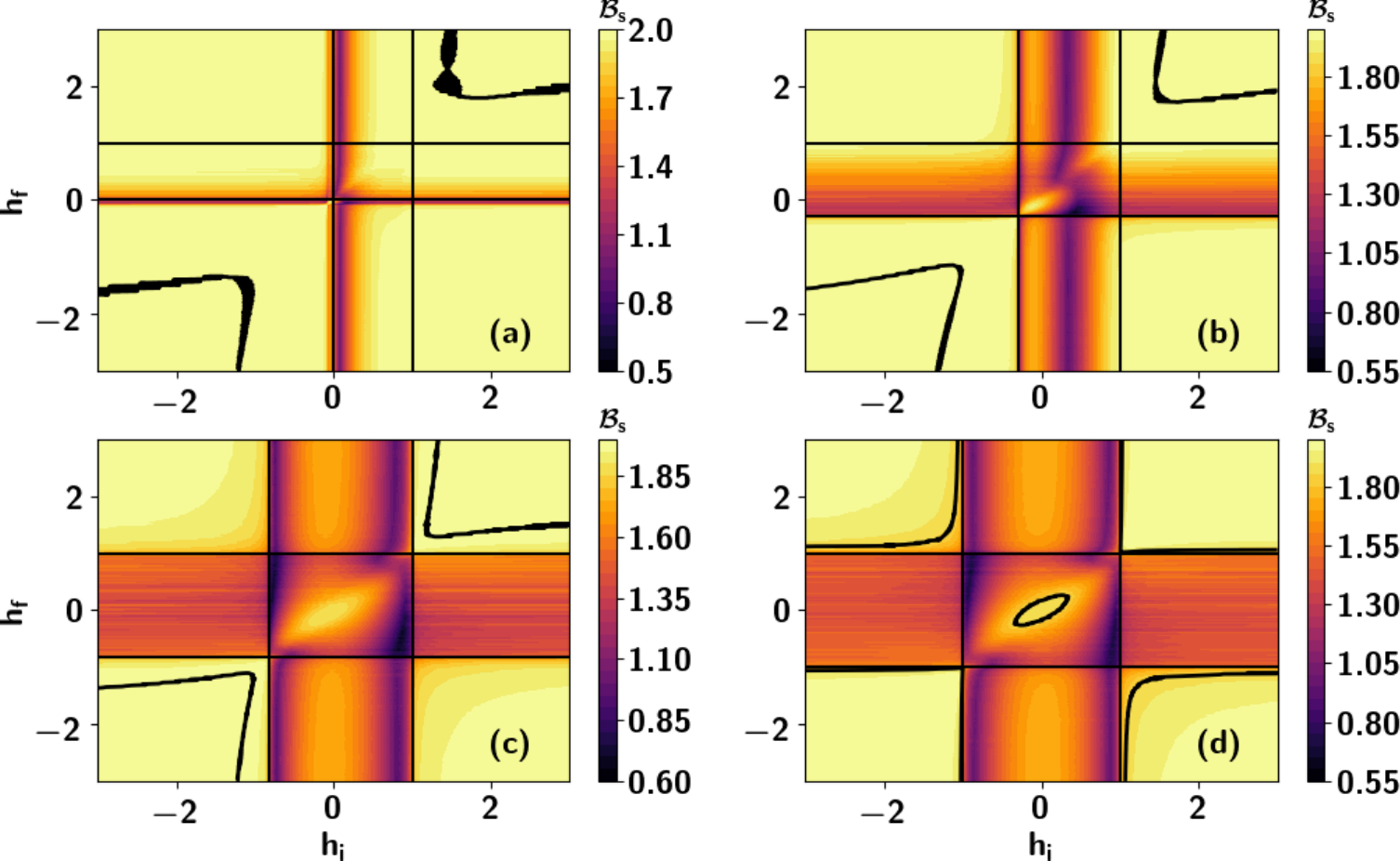}
    \caption{(Color online) {\bf Contour corresponding to the benchmarking value $\mathcal{B}_c$ in the parameter space $(h_i,h_f)$ of quantum XY-model with $\gamma = 0.2$.}
    The four panels correspond to the map plot of $\mathcal{B}_{s}$ for the long-range $XY$ models characterized by the fall-off rates, $\alpha$ -- (a) $\alpha = 0.9$, (b) $\alpha =1.5$, (c) $\alpha =3.5$, and (d) the nearest-neighbor limit (\(\alpha =10\)) by varying \(h_i\) (horizontal axis) and \(h_f\) (vertical axis).  The contours (solid  lines) defined by the benchmarking value $\mathcal{B}_c$ delineates the region associated with same-phase detection, and the enclosed area serves as a figure of merit for the efficiency of the proposed quantifier. All axes are dimensionless.}
    \label{Fig:h_quench_gamma_0.2}
\end{figure}

\subsubsection{Impact of Bell operator with  field quench}

The magnetic parameter space of the long-range $XY$ model is divided by the boundaries of $\{h_i, h_f\} \in \{-1+2^{1-\alpha}, 1\}$, which indicate the corresponding magnetically ordered and disordered phases. For quenches in the transverse field $h$, we establish the benchmark value $\mathcal{B}_c$, which remains robust across a wide range of fall-off rate $\alpha$ and anisotropy parameters $\gamma$ (see Table~\ref{Tab:Bc1-benchmark_h}). 

\begin{table}[h]
    \centering
    \begin{tabular}{|c|c|c|c|c|}
        \hline
        \diagbox{$\gamma$}{$\alpha$} & 0.9 & 1.5 & 3.5 & NN \\
        \hline
        0.2 & 1.9997 & 1.99843 & 1.98482 & 1.83393 \\
        0.8 & 1.99501 & 1.977 & 1.86674 & 1.71068 \\
        1.0 & 1.99254 & 1.96515 & 1.83578 & 1.71403 \\
        \hline
    \end{tabular}
    \caption{{\bf Benchmarking magnetic quench through Bell-based DQPT quantifier $\mathcal{B}_c$.} The table lists the threshold values of the steady-state Bell correlator, $\mathcal{B}_{s}(h_i, h_f, \alpha, \gamma)$, denoted as $\mathcal{B}_c$, for various combinations of $\alpha$ and $\gamma$. These thresholds indicate the critical values above which one can infer that the initial and final field strengths, $h_i$ and $h_f$, correspond to the same dynamical phase. Here, the ground state of the Hamiltonian $H(\alpha, \gamma, h_i)$ evolves under the quenching Hamiltonian $H(\alpha, \gamma, h_f)$, with $h_i$ and $h_f$ representing the initial and post-quench field strengths, respectively. Data are obtained for a system size of $N = 512$, by varying $(h_i, h_f)$  from $-3$ to $3$ in steps of $0.01$.}
    \label{Tab:Bc1-benchmark_h}
\end{table}

\begin{figure}
    \centering
    \includegraphics[width=\linewidth]{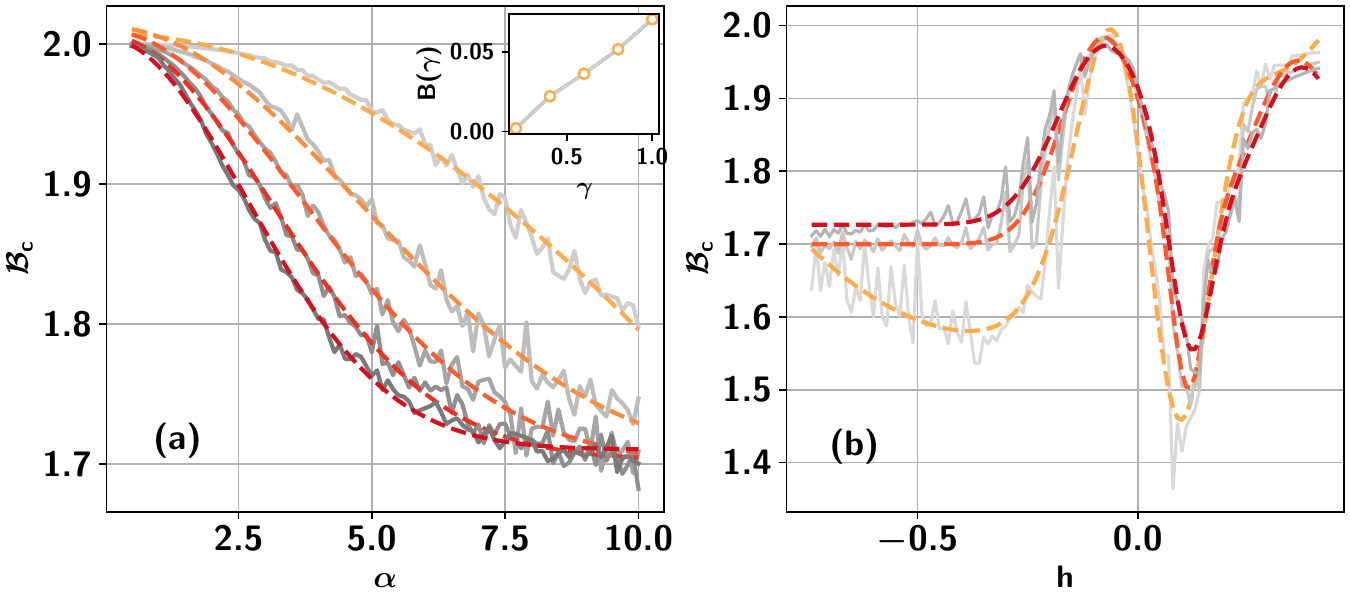}
    \caption{(Color online) {\bf  Critical benchmarking threshold as DQPT quantifier.}
    (a) Variation of $\mathcal{B}_c$ (solid line, ordinate) with $\alpha$ (abscissa) for  $\gamma = 0.2, 0.4, 0.6, 0.8$ and $1.0$ during magnetic field quenches, \((h_i, h_f)\). The dashed lines are the Gaussian fit, given in Eq. (\ref{eq:fieldqfit}). (Inset) Dependence of the inverse of variance of the fitted Gaussians on $\gamma$ upto a scale. 
    (b)  $\mathcal{B}_c$ (ordinate) vs $h$  (abscissa) within the range $[-0.75,0.41]$ for  $\gamma = 0.4, 0.8$ and $1.0$ during coupling quenches (\(\alpha_i, \alpha_f\)). In this case, the fitted dash curve is modeled after a tri-Gaussian function (see Eq. (\ref{eq:tri_gaussian_fit})).
    In both the cases, dark to light shades represent the decrease of \(\gamma\). 
    All the axes are dimensionless.}
    \label{Fig:Bc1_fit}
\end{figure}

\begin{figure}
    \centering
    \includegraphics[width=\linewidth]{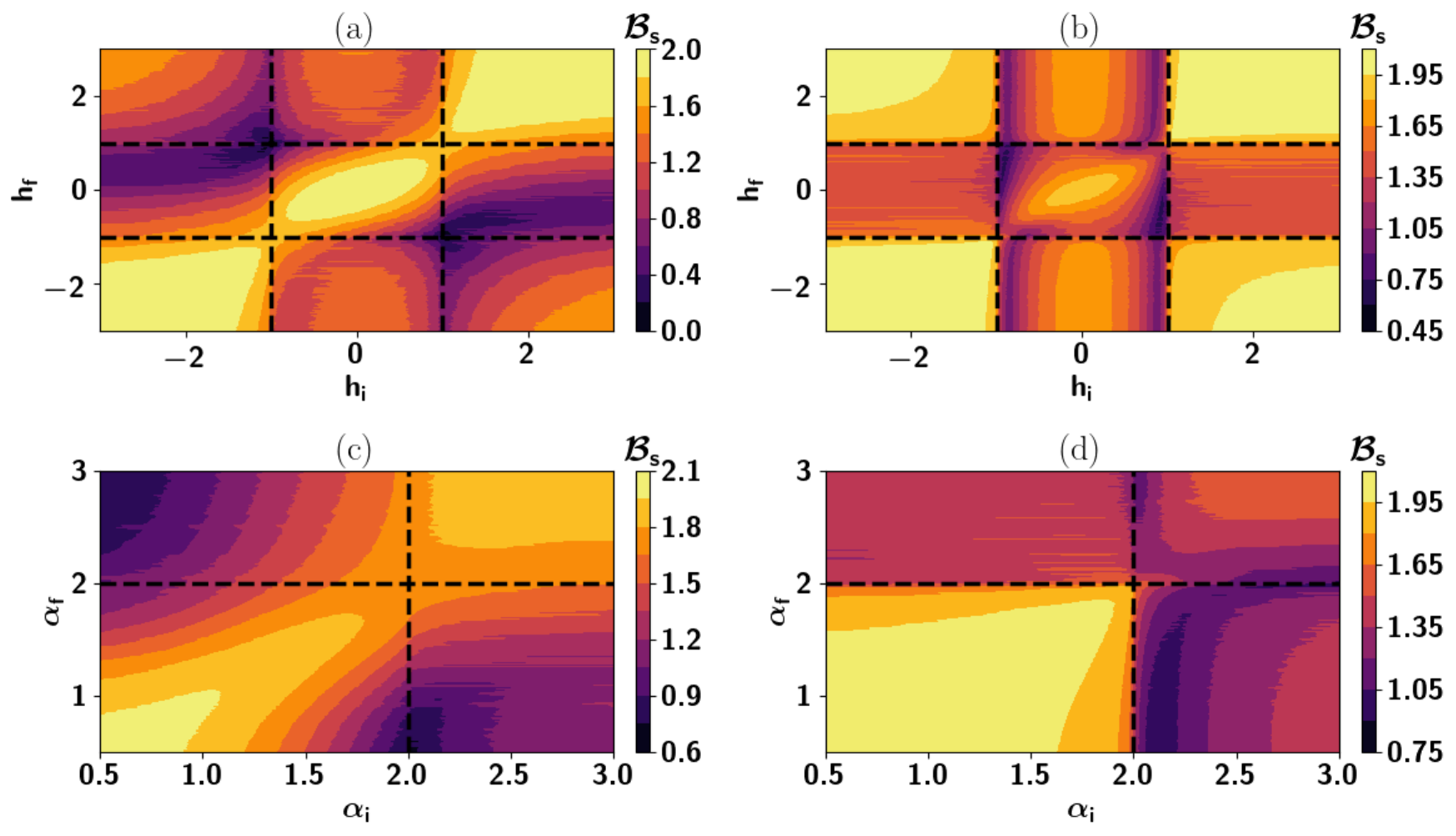}
    \caption{(Color online) {\bf Dependence of the sensitivity of the proposed quantifier on the anisotropy parameter $\gamma$.}
    The four panels correspond to the map plot of $\mathcal{B}_{s}$ for the $XY$ model characterized by the field quench  with \(\alpha =10\) ((a) and (b)) and the coupling quenches, having a fixed magnetic field, \(h =-0.5\)((c) and (d)). We compare the sharpness of the quantifier for different values of the anisotropy parameter -- $\gamma = 1.0$ ((a), (c)) and $0.2$ ((b), (d)). All axes are dimensionless.}
    \label{Fig:Sensitivity}
\end{figure}

The contours corresponding to these benchmark values (see Fig.~\ref{Fig:h_quench_gamma_0.2}) further delineate the parameter regions distinguishable by our proposed quantifier. We generate the values of $\mathcal{B}_c$ for a spin chain of length $N=512$ upto $\alpha=10$ in intervals of $\Delta \alpha = 0.1$ for fixed $\gamma$-values and plot them with respect to $\alpha$. The best fit for the trend followed by the set of data points $(\alpha, \mathcal{B}_c(\alpha,\gamma))$ appears to be Gaussian in nature for a fixed anisotropy $\gamma$, given by
\begin{equation}
\mathcal{B}_c(\alpha,\gamma) = A\exp[-
B (\gamma) \alpha^2]+C, 
\label{eq:fieldqfit}
\end{equation}
where $A,~ B$ and $C$ are functions of parameters of the Hamiltonian. Moreover, we observe an interesting pattern for different $\gamma$: the Gaussian fit becomes narrower with increasing $\gamma$, pointing to a possible $\gamma$-dependence of the standard deviations of  data for \(\mathcal{B}_c(\alpha,\gamma)\) (see Fig. \ref{Fig:Bc1_fit} (a)).
Interestingly, the sensitivity of \(\mathcal{B}_c\) to variations in the anisotropy parameter \(\gamma\) is rather remarkable, since it directly dictates sharpness of the quantifier.  To capture this,  we compare \(\mathcal{B}_s\) for two values \(\gamma\) in Fig. \ref{Fig:Sensitivity}(a) and (b) demonstrating that as \(\gamma\) decreases, \(\mathcal{B}_s\) becomes more active to detect quantum phases dynamically.

\begin{figure}
    \centering
    \includegraphics[width=\linewidth]{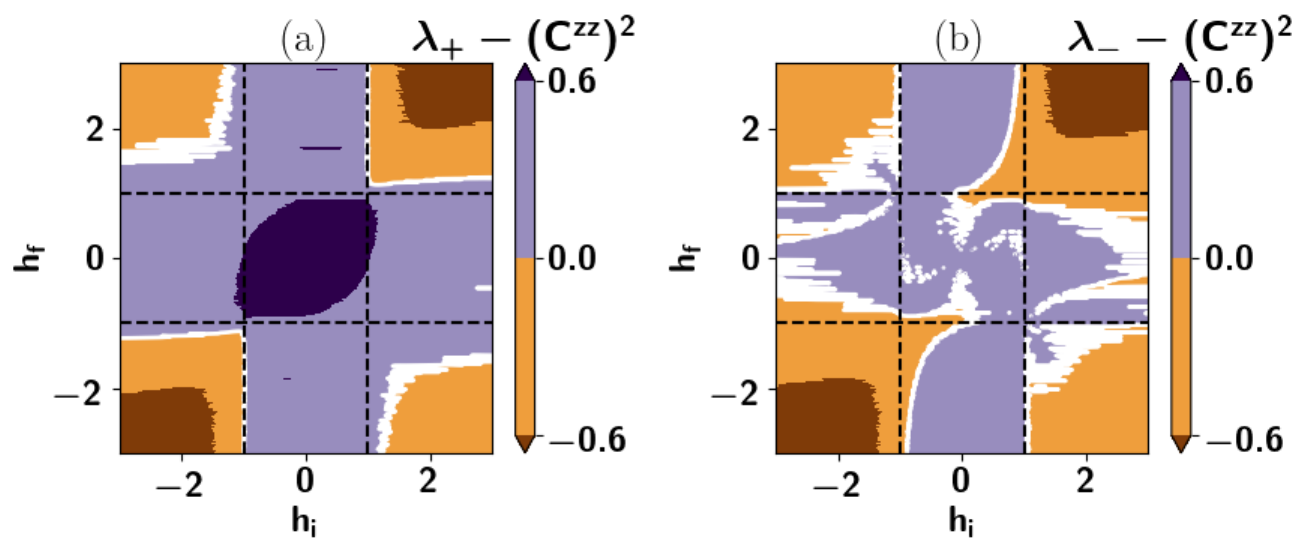}
    \caption{(Color online) {\bf Competition between the eigenvalues of the steady-state Bell correlator.}
    (a) Map plot of $\lambda_+ - (C^{zz})^2$ with (\(h_i, h_f\))-pair (horizontal and vertical axes) for the NN transverse Ising model. The competition between \(\lambda_{-}\)
    and $(C^{zz})^2$ across the parameter space, with purple regions indicating dominance of $\lambda_-$ and orange regions indicating dominance of $(C^{zz})^2$. (b) $\lambda_{-} - (C^{zz})^2$ with (\(h_i, h_f\))-pair. 
     Here, purple and orange regions correspond to the dominance of $\lambda_+$ and $(C^{zz})^2$, respectively. White-colored regions mark the zero-valued parts. All axes are dimensionless.}
    \label{Fig:hBelleigdiff}
\end{figure}

{\it Microscopic origin of the steady-state values.} As emphasized earlier in Eq.~(\ref{Eq.Belleigs}), the Bell correlation of the reduced two-party density matrix obtained from the unitary evolution of the initial ground state of the $XY$ model is obtained due to the competition between  two eigenvalues of $\mathcal{M}$, namely $\{(C^{zz})^2, \lambda_-\}$. Let us argue this with the NN Ising model. From Fig.~\ref{Fig:hBelleigdiff}, it is evident that the steady-state Bell value $\mathcal{B}_s$ resulting from a same-phase quench initiated in the paramagnetic phase is typically dominated by $(C^{zz})^2$, whereas in quenches from the ferromagnetic phase or in quenches to the ferromagnetic phase, its behavior is dictated by $\lambda_-$. Remarkably, in the case of same phase quenches to the ferromagnetic phase, the steady-state Bell value is largely influenced by the eigenvalue $\lambda_+$, as clearly shown in Fig.~\ref{Fig:hBelleigdiff}(a).

{\it Dynamical sensitivity near criticality.} Our numerical results also reveal a more subtle dynamical behavior associated with $(C^{zz})^2$ and \(\mathcal{B}_s\). Firstly, $(C^{zz})^2$ dominates both $\lambda_-$ and $\lambda_+$ in same-phase quenches within the paramagnetic region (see Fig.~\ref{Fig:hBelleigdiff}). Interestingly, its contribution diminishes significantly when the initial and final magnetic phases are the same, although it crosses the critical line \cite{Stav_PRB_2020}. 
 This reduction highlights the dynamical sensitivity of $(C^{zz})^2$ and \(\mathcal{B}_s\) to criticality and indicates that equilibrium phase transitions leave a residual fingerprint in non-equilibrium correlators.

\begin{table}[h]
    \centering
    \scriptsize
    \setlength{\tabcolsep}{2.1pt} 
    \begin{tabular}{|c|cccc|cccc|cccc|}
        \hline
        & \multicolumn{4}{c|}{$\eta_\mathcal{B}$ } 
        & \multicolumn{4}{c|}{$\eta_\mathcal{E}$ } 
        & \multicolumn{4}{c|}{$\eta_C$ } \\
        \hline
        \diagbox{$\gamma$}{$\alpha$} & 0.9 & 1.5 & 3.5 & NN & 0.9 & 1.5 & 3.5 & NN & 0.9 & 1.5 & 3.5 & NN \\
        \hline
        0.2 & 0.18 & 0.20 & 0.29 & 0.75 & 0 & 0 & 0 & 0 & 0.18 & 0.23 & 0.34 & 0.75 \\
        0.8 & 0.19 & 0.20 & 0.30 & 0.41 & 0.0007 & 0.005 & 0 & 0.034 & 0.18 & 0.21 & 0.29 & 0.45 \\
        1.0 & 0.18 & 0.20 & 0.31 & 0.41 & 0.0004 & 0.007 & 0 & 0.045 & 0.18 & 0.22 & 0.28 & 0.38 \\
        \hline
    \end{tabular}
    \caption{{\bf Comparative efficiencies under magnetic quenches.} 
    Efficiency values of three quantifiers, $\eta_\mathcal{B}$ (Bell-based), $\eta_\mathcal{E}$, (entanglement-based), and $\eta_C$ ($C^{zz}$-based) for different $(\alpha,\gamma)$ combinations.
    }
    \label{Tab:combined-efficiency_h_Bc1}
\end{table}

The behavior of the Bell-based efficiency $\eta_\mathcal{B}$ displays a revealing pattern (see Table~\ref{Tab:combined-efficiency_h_Bc1}). Across all anisotropy values, the efficiency remains finite and exhibits a clear enhancement as the interaction range becomes shorter, attaining its maximum in the nearest-neighbor  limit. For instance, at $\gamma=0.2$, $\eta_\mathcal{B}$ grows from approximately $0.18$–$0.29$ in the long-range regime to $0.75$ in the NN case, signifying a robust detection capability in the strongly local limit. Similar trends persist for $\gamma=0.8$ and $\gamma=1.0$, where $\eta_\mathcal{B}$ maintains values between $0.18$ and $0.41$, indicating that the Bell correlator consistently preserves its ability to identify intra-phase quenches even when the system approaches $\gamma=1$.

\subsubsection{Trends of Bell operator under coupling quench}
\label{sec:Belloperatoralphaquench}

When the system is quenched with respect to the long-range interaction strength, $\alpha$, in the parameter space $(\alpha, h)$ while  the external magnetic field $h$ remains fixed, the magnetic boundary located at $k=\pi$ in the momentum space undergoes a shift. Consequently, for a fixed field strength within the interval $[-1+2^{1-\alpha_i}, -1+2^{1-\alpha_f}]$, the system may cross the magnetic boundary corresponding to $k=\pi$ for certain combinations of $(\alpha_i, \alpha_f)$. As a result, distinct magnetic phases emerge in the parameter space of $\alpha$. In our analysis,  the minimum and maximum values for $\alpha$  are taken to be $0.5$ and $3.0$, respectively. It implies that $h$ must lie in the range $(-0.75,0.414)$ to observe the magnetic boundary in the $\alpha$-parameter space. In this section, we demonstrate that the proposed quantifier $\mathcal{B}_c$ effectively captures these magnetic phases, and its sensitivity is governed by the anisotropy parameter $\gamma$ of the quantum $XY$ model. 

\begin{figure}
    \centering
    \includegraphics[width=\linewidth]{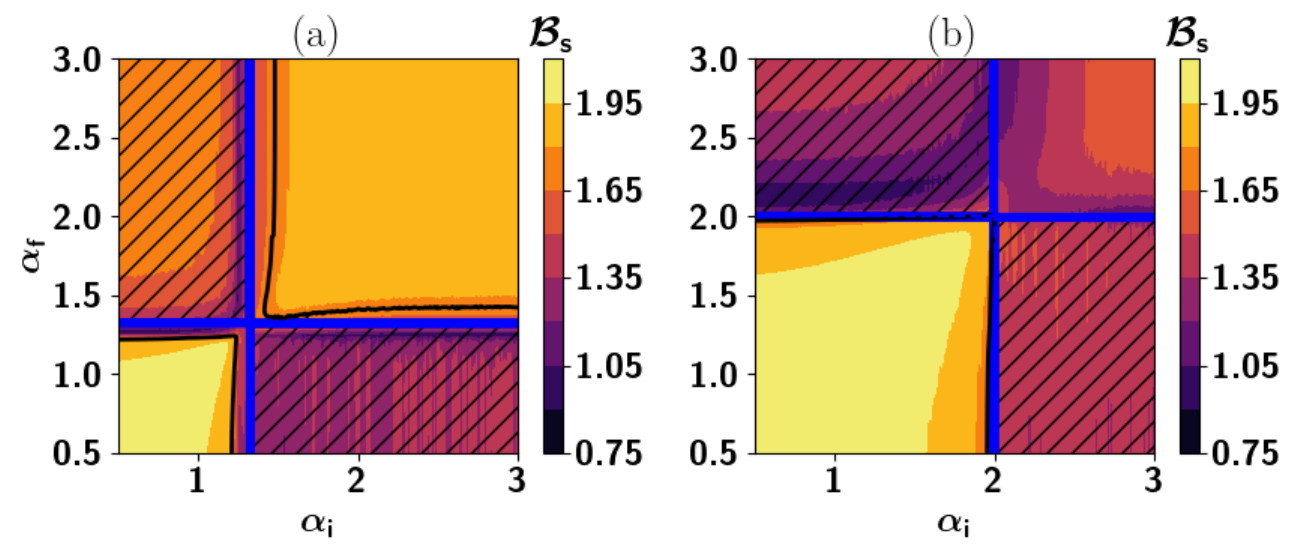}
    \caption{(Color online) {\bf Detection of magnetic boundary through coupling quench. } Projected contour plot of $\mathbf{\mathcal{B}_{s}}$ in the quench parameter pair, $(\alpha_i,\alpha_f)$ (vertical-horizontal axes pair), for $\gamma = 0.2$ with
    (a) $h=-0.2$, and (b) $h=-0.5$. Blue lines differentiate magnetic phases during the coupling quench, and the hatched regions define the quenches when the system crosses a magnetic boundary. The black contour line is the benchmarking contour defined by $\mathcal{B}_c$. All the axes are dimensionless.    } 
    \label{Fig:Bc1_with_alpha}
\end{figure}

For a fixed external field $h$, the system crosses the magnetic boundary when $h=-1+2^{1-\alpha_c}$, from which we can infer these new magnetic boundaries emerge at $\alpha_c=1-\log_2(1+h)$. We observe that during $\alpha$-quench, $\mathcal{B}_{s}$ values are significantly higher when the system does not cross a magnetic boundary compared to the case with inter-phase quench. Similar to the field quench, we again find that the boundary obtained from the steady-state Bell correlator is increasingly sharper as the anisotropy, \(\gamma\), decreases, indicating that $\gamma$ governs the sensitivity of the quantifier \(\mathcal{B}_c\) (See Fig.~\ref{Fig:Sensitivity}(c) and (d)).

\begin{figure}
    \centering
    \includegraphics[width=\linewidth]{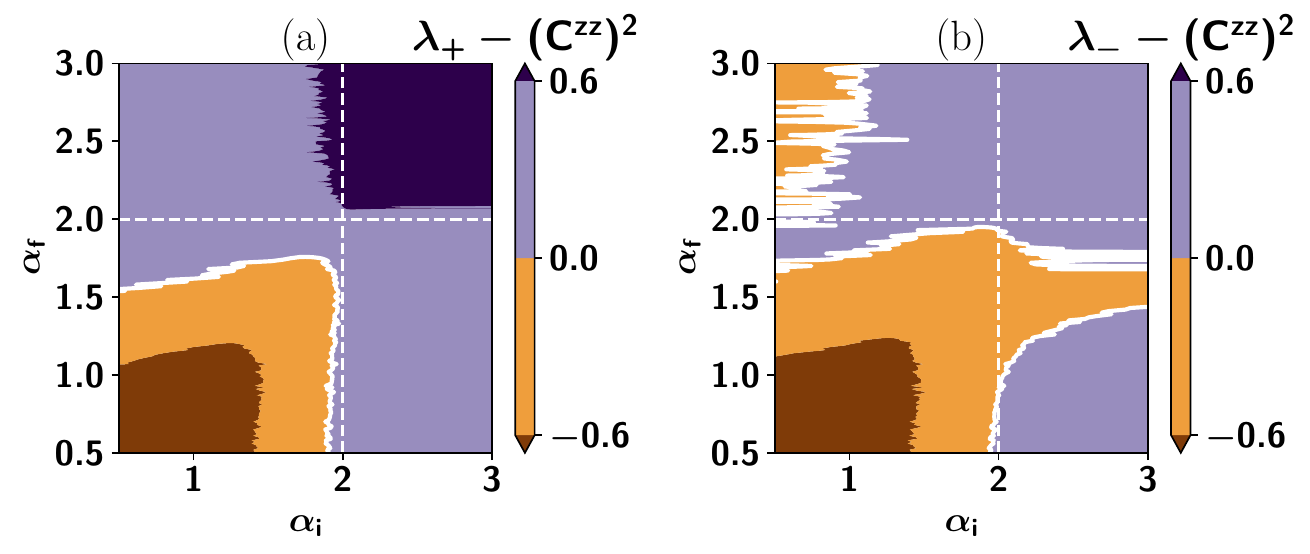}
    \caption{(Color online) {\bf Competition between eigenvalues of the steady-state Bell correlator in the $(\alpha_i, \alpha_f)$ parameter space of the variable-range Ising model during coupling quench.}
    Here $h = -0.5$. All other specifications and interpretations are the same as in Fig. \ref{Fig:hBelleigdiff}. 
    All axes are dimensionless.}
    \label{Fig:aaBelleigdiff_0.2}
\end{figure}

{\it Microscopic origin of the steady state values.} A detailed examination of the steady-state Bell correlator reveals a remarkable feature: its values are predominantly governed by $(C^{zz})^2$ when the coupling parameter is quenched such that the system remains within the same magnetic phase and in the non-local interaction regime. Notably, this is the region where the Bell-based quantifier $\mathcal{B}_c$ effectively discriminates between distinct behaviors, indicating that the enhanced values of $\mathcal{B}_{s}$ arise from the dominance of $(C^{zz})^2$. In contrast, the $\mathcal{B}_{s}$ values corresponding to other regions of the quench parameter space $(\alpha_i, \alpha_f)$ are primarily determined by $\lambda_-$, exhibiting lower values and enabling us to determine the phase structure by comparing the values of the steady-state Bell correlator.

During the coupling quench, the proposed DQPT quantifier $\mathcal{B}_c$ as a function of the external field $h$ (see Table.~\ref{Tab:Bc1-benchmark_hDetection_aa}), for a fixed value of the anisotropy parameter $\gamma$, shows a nontrivial and oscillatory profile. It can be quantitatively captured by a tri-Gaussian fitting function as
\begin{equation}
    \mathcal{B}_c(h) = \sum_{i=1}^{3} A_i \exp\!\left[-\frac{(h - \mu_i)^2}{2\sigma_i^2}\right],
    \label{eq:tri_gaussian_fit}
\end{equation}
where $\{A_i, \mu_i, \sigma_i\}$ denote the amplitude, central position, and width of each Gaussian component, respectively. As shown in Fig.~\ref{Fig:Bc1_fit}(b), the gray curve represents the numerically obtained steady-state values, while the superimposed red and orange dashed lines correspond to the tri-Gaussian fits for two representative parameter sets. The extracted fitting parameters exhibit systematic and physically interpretable dependencies on the system parameters $(\alpha, \gamma)$. Specifically, the central positions $\mu_i(\alpha)$ shift monotonically with \(\alpha\), reflecting the displacement of the magnetic boundary at $k=\pi$ as governed by $h = -1 + 2^{1-\alpha}$.

{\it Efficacy of the Bell correlator.} A closer inspection of the efficiency values corresponding to $\mathcal{B}_c$ in Table~\ref{Tab:combined-efficiency_hDetection_aa_Bc1} reveals a remarkably consistent trend: the Bell-based quantifier retains a high degree of efficiency across distinct parameter regimes, thereby establishing its robustness as a dynamical indicator of equilibrium phase structure. For example, for weak anisotropy ($\gamma = 0.2$), the efficiency turns out to be moderate to high, viz. $0.85$, $0.67$, and $0.93$ for  field values, \(-0.2\), \(-0.5\), and  \(-0.7\) respectively. This signifies that the entire phase diagram corresponding to same-phase quenches remains distinguishable under Bell diagnostics. Even as the anisotropy increases to $\gamma = 0.8$ and $\gamma = 1.0$, the efficiency remains more than \(50\%\) ($\sim 0.5$–$0.75$), underscoring that $\mathcal{B}_c$ retains its discriminative capacity irrespective of the anisotropy parameters.

\begin{table}[h] 
\centering 
\begin{tabular}{|c|c|c|c|} 
\hline
\diagbox{$\gamma$}{$h$} & -0.2 & -0.5 & -0.7\\ 
\hline
0.2 & 1.72533 & 1.6562 & 1.75671\\
0.8 & 1.89374 & 1.71845 & 1.72372\\
1.0 & 1.90389 & 1.74951 & 1.72674\\
\hline
\end{tabular}
\caption{{\bf Benchmarking $\mathcal{B}_c$ during detection of magnetic boundaries in coupling quench.}
Threshold values of the steady-state Bell correlator $\mathcal{B}_s(\alpha_i,\alpha_f,h,\gamma)$ are shown for $\mathcal{B}_c$ (same-phase criterion). $\mathcal{B}_c^{(2)}$ (cross-phase criterion) does not exist in this case. 
Data correspond to $N=512$ with $(\alpha_i,\alpha_f)$ being chosen from $0.5$ to $3.0$ in steps of $0.01$.} 
\label{Tab:Bc1-benchmark_hDetection_aa}
\end{table}

\section{Correlators and Bipartite  entanglement fail to detect DQPT}
\label{sec:bell_vs_ent_and_correlator}

Entanglement, often considered a defining feature of quantum systems, serves as a fundamental diagnostic for exploring various quantum correlations in many-body systems. Nevertheless, despite its significance, entanglement does not universally succeed in capturing the onset of quantum criticality. Previous studies have already demonstrated that, in equilibrium scenarios, the magnitude of bipartite entanglement is typically weaker than Bell correlations (see also \cite{Stav_PRB_2020}). Extending this premise, our analysis establishes that such limitations persist even in the dynamical regimes: specifically, the steady-state entanglement of a bipartite reduced state following a quench is found to be an unreliable indicator of detecting quench across equilibrium critical lines. 

\textbf{Observation 3.} \textit{Bipartite entanglement and the steady-state classical correlators except $C^{zz}$ are not as efficient as the Bell correlator for the dynamical detection of phases in the long-range quantum $XY$ model.}

\begin{figure}[h]
    \centering
    \includegraphics[width=\linewidth]{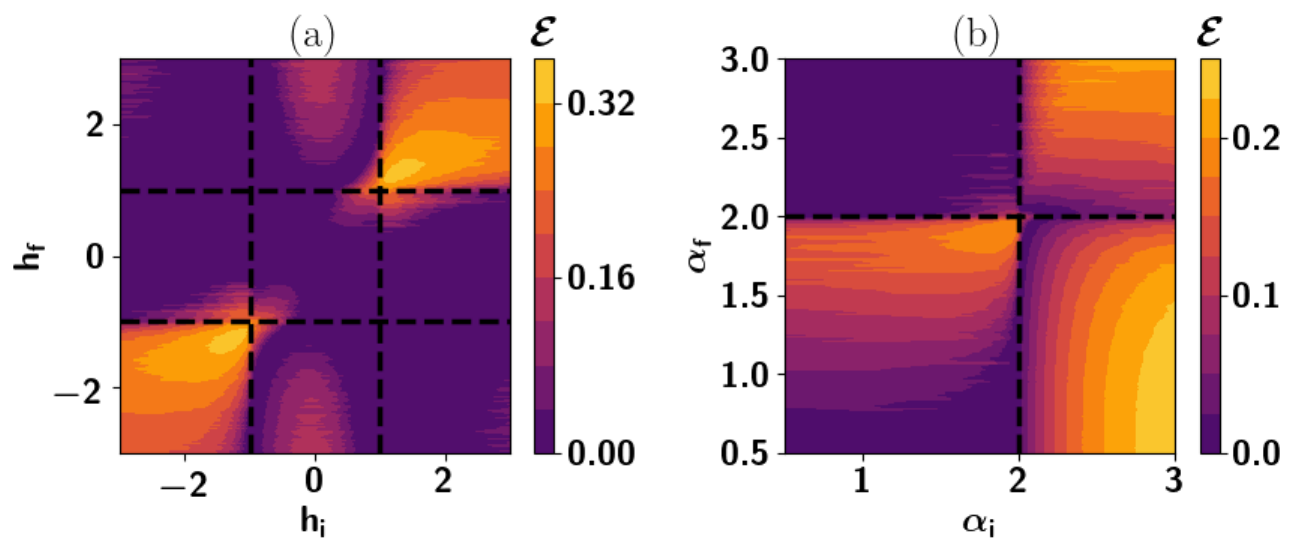}
    \caption{(Color online) {\bf Saturated bipartite entanglement.}
    (a) Map plot of nearest-neighbor entanglement of the quantum transverse Ising model in the \((h_i, h_f)\)-plane. (b) The same in the \((\alpha_i, \alpha_f)\)-plane with $h=-0.5$. Here $\gamma = 0.2$.  The horizontal and vertical dashed black lines represent the magnetic phase boundaries. It is evident that the bipartite entanglement of the evolved state cannot identify the equilibrium phases. All axes are dimensionless. }
    \label{Fig:Entanglement_vs_Bell_gamma_0.2}
\end{figure}

\begin{figure}[h]
    \centering
    \includegraphics[width=\linewidth]{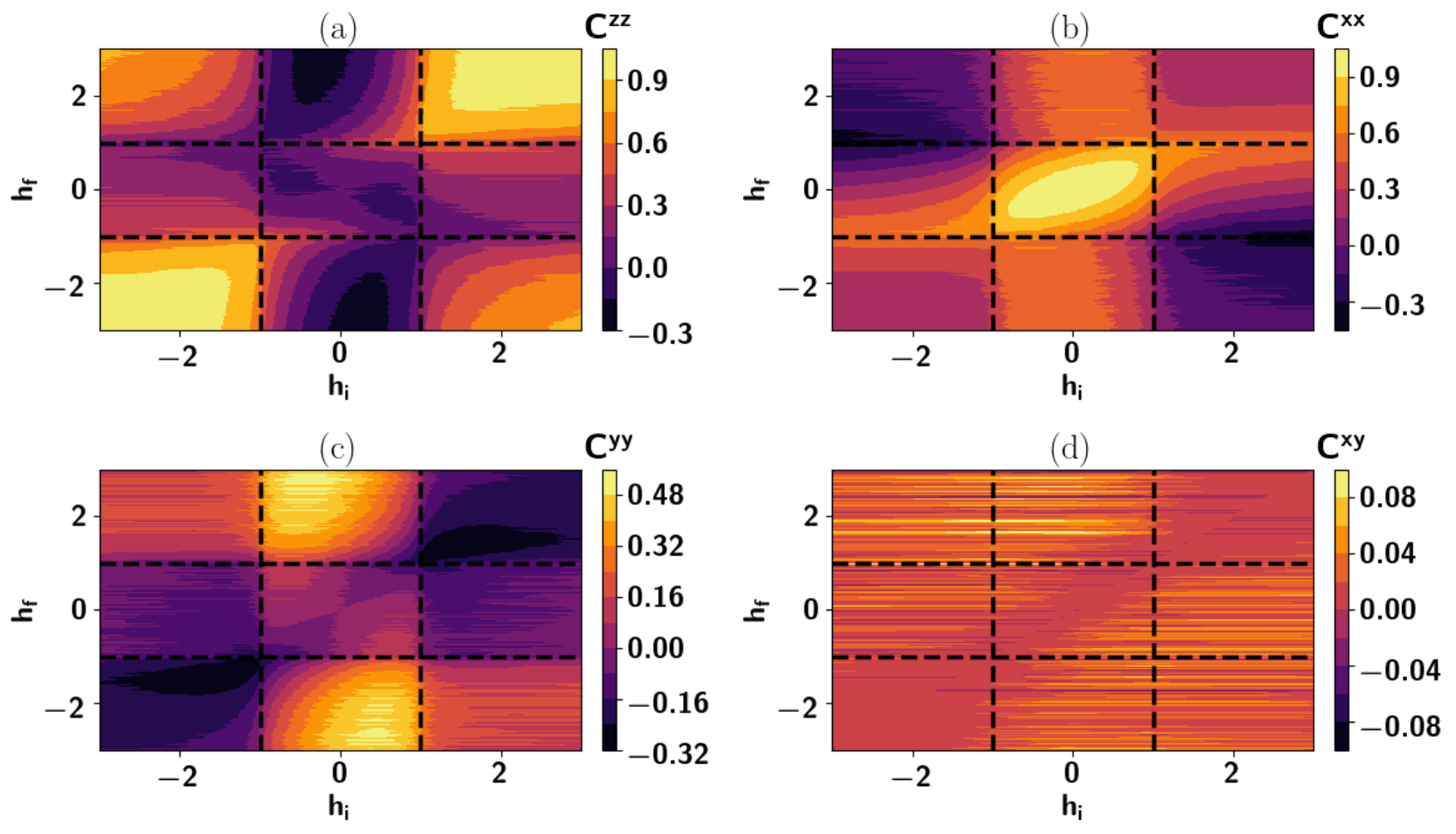}
    \caption{(Color online) {\bf Comparison between different classical correlators.}
    The four panels correspond to the saturation value of the classical correlators of the nearest-neighbor transverse Ising model during magnetic quench, (\(h_i, h_f\)).  (a) $C^{zz}$, (b) $C^{xx}$, (c) $C^{yy}$ and (d) $C^{xy}$.  The horizontal and the vertical lines are the quantum critical lines, $h_i = \pm 1$, and $h_f = \pm 1$. All axes are dimensionless. Here $\gamma = 0.2$. Clearly, classical correlators except \(C_zz\) fail to determine DQPT. All the axes are dimensionless. }
    \label{Fig:classical_vs_bell_gamma_0.2}
\end{figure}

To substantiate this claim, we analyze the performance of steady-state bipartite entanglement, quantified via the logarithmic negativity \cite{plenio2005, vidal2002} of the reduced state $\rho_{12}(t)$, denoted by $\mathcal{E}$ \cite{log_neg}. In parallel, we compute the steady-state values of all classical correlators ($C^{xx}, C^{yy}, C^{zz}, C^{xy}$) for representative field quenches, as depicted in Fig.~\ref{Fig:classical_vs_bell_gamma_0.2}; analogous observations hold for coupling quenches.  
While $C^{xx}, C^{yy}$, and $C^{xy}$, including $\mathcal{E}$, remain largely insensitive to phase boundaries, $C^{zz}$ (as discussed before) exhibits a modest degree of differentiability, capturing certain critical features albeit less reliably than the Bell correlator. By contrast, the Bell correlator $\mathcal{B}_{s}$ manifests a pronounced, phase-dependent response, effectively resolving the critical boundaries across all quench regimes (compare Figs. \ref{Fig:classical_vs_bell_gamma_0.2} and \ref{Fig:Bell_satVSavg} (a)). This distinct behavior highlights the superior sensitivity of Bell correlations compared to  two-point stand-alone classical observables.  

Motivated by these qualitative observations, we undertake an efficacy study among $\mathcal{B}_{s}$, $C^{zz}$, and $\mathcal{E}$, to quantitatively assess their respective abilities to detect dynamical quantum phase transitions.  The efficacy parameters are indicated as  $\eta_\mathcal{E}$ (entanglement-based), and $\eta_C$ ($C^{zz}$ based), which are defined analogously to $\eta_\mathcal{B}$ (Bell-based).  This approach allows us to systematically elucidate the relative merits of entanglement, classical correlations, and Bell-based measures in capturing the dynamical signatures of criticality.

\textit{Field-quench analysis.}
In the case of field quenches, the correlator-based quantifier, \(C^{zz}\) performs remarkably well, showing sensitivity comparable to that of the Bell correlator although other two-site classical correlators fail to perform. Specifically, $\eta_C \in (0.18, 0.75)$ for \(C^{zz}\) across all $(\gamma,\alpha)$ combinations,
 combinations, indicating consistent responsiveness as the interaction range evolves from long-range to the nearest-neighbour limit. This robustness underscores its comparable efficiency to the Bell correlator in detecting dynamical signatures under field quenches. 

In contrast, $\eta_\mathcal{E}$ displays negligible response, with efficiency values effectively vanishing for weak and intermediate interactions and increasing only marginally (up to $0.045$) in the NN regime (see Table. \ref{Tab:combined-efficiency_h_Bc1}).

Interestingly, the relative stability of $\eta_\mathcal{B}$ across $\gamma$ contrasts sharply with the sensitivity of $\eta_\mathcal{E}$, whose efficiency fluctuates without systematic dependence on either $\gamma$ or $\alpha$. This further supports the notion that bipartite entanglement, though sufficient to characterize certain static transitions, lacks the dynamical resolution necessary to capture DQPTs emerging from long-range quench processes.

\textit{Coupling quench.} We next turn to the long-range coupling quenches, summarized in Table~\ref{Tab:combined-efficiency_hDetection_aa_Bc1}. Across all $(\gamma,h)$ pairs, $\eta_\mathcal{B}$ again outperforms $\eta_\mathcal{E}$ by a wide margin, reaching efficiencies above $0.8$ for weak anisotropy ($\gamma=0.2$) and maintaining values between $0.5$–$0.7$ for $\gamma \ge 0.8$. In contrast, $\eta_E$ barely exceeds $0.1$ even in favorable regions. Surprisingly, we emphasize the overwhelmingly high value of the efficacy of $\eta_\mathcal{B}$ as compared to $\eta_C$ throughout all values of $\gamma$ and $h$ unlike the field quench. Precisely, the Bell correlator exhibits outstanding performance at low magnetic field strengths compared two-site \(C^{zz}\), although its efficiency becomes comparable with  \(C^{zz}\) as the field strength increases.   

We find that, although specific classical correlators such as $C^{zz}$ reflect qualitative shifts near critical points, 
it is still less efficient than $\mathcal{B}_{s}$, especially in the coupling quench scenario ($h=-0.2$). This suggests that the Bell signal encodes non-trivial quantum information that is not reducible to classical two-point structures.

\begin{table}
    \centering
    \scriptsize
    \setlength{\tabcolsep}{2.1pt} 
    \begin{tabular}{|c|ccc|ccc|ccc|}
        \hline
        & \multicolumn{3}{c|}{$\eta_\mathcal{B}$ } 
        & \multicolumn{3}{c|}{$\eta_\mathcal{E}$ } 
        & \multicolumn{3}{c|}{$\eta_C$ }
        \\
        \hline
        \diagbox{$\gamma$}{$h$} & $-0.2$ & $-0.5$ & $-0.7$ & $-0.2$ & $-0.5$ & $-0.7$ & $-0.2$ & $-0.5$ & $-0.7$ \\
        \hline
        0.2 & 0.85 & 0.67 & 0.93 & 0 & 0 & 0.20 & 0.16 & 0.68 & 0.94 \\
        0.8 & 0.76 & 0.64 & 0.53 & 0.0280 & 0.0698 & 0.13 & 0.15 & 0.55 & 0.67 \\
        1.0 & 0.74 & 0.58 & 0.47 & 0.0244 & 0.0772 & 0.14 & 0.14 & 0.51 & 0.64 \\
        \hline
    \end{tabular}
    \caption{{\bf Efficiencies under coupling quenches.} 
Instead of \((h_i, h_f)\) in Table \ref{Tab:combined-efficiency_h_Bc1}, these data are obtained with \((\alpha_i, \alpha_f)\)-pair. All other specifications are the same as in Table. \ref{Tab:combined-efficiency_h_Bc1}.}
    \label{Tab:combined-efficiency_hDetection_aa_Bc1}
\end{table}

\section{Operational Perspective and Experimental Context}
\label{sec:exper}

 Prior proposals have employed quasi-local string operators acting on approximately seven spins to distinguish phases. However, implementing these operators in practice is challenging, as they generally require comparing the system before and after the evolution, leading to substantial tomography requirements and increased measurement complexity \cite{bandyopadhya_prl_2021, souvik_experiment}. This naturally motivates the search for phase-diagnostic tools that (i) act locally, (ii) equilibrate rapidly following a quench, and (iii) avoid global information requirements while still retaining sensitivity to the underlying phase structure.

The Bell correlator introduced in this work meets all three criteria. First, because the observable acts only on a pair of neighboring spins, its steady-state value is reached within experimentally accessible time windows (see Fig. \ref{Fig:BelldropObs}), even in systems where long-range interactions lead to nontrivial dynamical behavior. Although the steady-state value plays the primary role in our analysis, it is noteworthy that the relaxation time is relatively short, thereby lowering the practical overhead on temporal evolution and mitigating decoherence concerns.

Second, previous dynamical studies aimed at characterizing quenched phases have largely focused on the Ising limit $(\gamma = 1)$. By contrast, the present investigation explicitly explores a range of anisotropy values and establishes that the Bell-based indicator remains sensitive to both intra- and inter-phase quenches beyond the Ising case. Thus, the Bell correlator not only captures the dynamical phase transition structure but also extends its relevance across a broader parameter regime, providing a versatile probe for interacting spin systems with tunable anisotropy.

Third, although the dynamical detection of equilibrium phases in long-range interacting systems has been considered earlier, existing analyses predominantly address models belonging to the Ising universality class. Here, we examine an anisotropic long-range model and demonstrate that the Bell correlator continues to differentiate equilibrium phases even in this more general scenario. This observation is particularly relevant in light of recent advances in trapped-ion and Rydberg-atom platforms, where controllable long-range interactions and local measurements are simultaneously available ~\cite{rydberg_review_experiments, quantum_gas_microscope_review}.

Taken together, these observations underscore the utility of the Bell correlator as a practical and conceptually appealing diagnostic tool for many-body phases in both equilibrium and nonequilibrium settings. Importantly, the interpretation here does not rely on demonstrating a violation of Bell nonlocality; rather, the relevant information emerges from how the correlator responds to changes in the system’s spectral and dynamical landscape. In this sense, the study highlights a broader principle: quantum-information-inspired observables can serve as sensitive markers of many-body structure even when employed within purely local measurement paradigms. 

\section{Conclusion}
\label{sec:conclu}
Quantum phase transitions, occurring at zero temperature through variations in system parameters of many-body systems, reveal emergent states of matter with no classical counterparts. It has been established that dynamical quantifiers such as the Loschmidt echo, and multipartite entanglement can identify quantum critical points through inter- and intra-phase quenches, phenomena known as dynamical quantum phase transitions (DQPT). Experimentally, however, detecting DQPTs remains a challenge, as conventional diagnostics typically demand global access to the system. Therefore, identifying experimentally accessible local or few-body observables capable of signaling DQPTs is highly desirable.

We proposed a two-body Bell Clauser-Horne-Shimony-Holt (CHSH) correlator as an efficient tool to identify quantum critical points during  dynamics. Specifically, we investigate a class of paradigmatic long-range 
$XY$ models under a magnetic field, where the interaction strength decays as a power law. The system is initially prepared in the ground state of the Hamiltonian and then driven out of equilibrium by varying two parameters, the magnetic field strength and the interaction fall-off rate, referred to as field and coupling quenches, respectively. In both cases, we found that the long-time (steady-state) value of the Bell correlator, termed the saturated Bell correlator, effectively indicates whether the system crosses a critical point during the quench. In particular, when the initial and quenched Hamiltonians belong to the same phase, the saturated Bell correlator attains a high value, whereas it decreases significantly when the quench drives the system across different phases.

Building on these qualitative observations, we introduced two key quantities,  the critical benchmarking threshold of the saturated Bell correlator and its corresponding efficiency. Our analysis revealed that the critical benchmarking values of the saturated Bell correlator follow Gaussian-like laws for both field and coupling quenches. Moreover, we demonstrated that the Bell correlator maintains a consistently high efficiency, determined by its critical threshold, regardless of the specific quenching parameters. In contrast, entanglement and other classical correlators exhibit significantly lower efficiency under similar conditions, underscoring the clear advantage of using the Bell correlator in many-body systems. Notably, the Bell correlator can be directly measured in experiments without requiring full-state tomography, further enhancing its practical utility. In conclusion, our study establishes that the steady-state two-body CHSH Bell correlator serves as a powerful and efficient tool for identifying magnetic phases in integrable systems such as the long-range extended quantum $XY$ chain under dynamical evolution. It surpasses conventional DQPT quantifiers, including the rate function and entanglement, and emerges as the physically motivated and experimentally accessible correlator.

This study underscores the  significance of the Bell correlator, extending its role beyond foundational investigations of locality and realism violations. Further, it  offers a powerful framework to explore, characterize, and understand the complex non-equilibrium behavior  of many-body quantum systems.

\acknowledgements

 We acknowledge the use of \href{https://github.com/titaschanda/QIClib}{QIClib} -- a modern C++ library for general purpose quantum information processing and quantum computing (\url{https://titaschanda.github.io/QIClib}), and the cluster computing facility at the Harish-Chandra Research Institute. This research was supported in part by the INFOSYS scholarship for senior students. A.M. acknowledges funding support for Chanakya - PhD fellowship from the National Mission on Interdisciplinary Cyber Physical Systems, of the Department of Science and Technology, Govt. of India, through the I-HUB Quantum Technology Foundation. LGCL is funded by the European Union. Views and opinions expressed are, however, those of the author(s) only and do not necessarily reflect those of the European Union or the European Commission. Neither the European Union nor the granting authority can be held responsible for them. This project has received funding from the European Union’s Horizon Europe research and innovation programme under grant agreement No 101080086 NeQST. In addition, this project has been funded by the Caritro Foundation. This work was supported by the Provincia Autonoma di Trento, and Q@TN, the joint lab between the University of Trento, FBK—Fondazione Bruno Kessler, INFN—National Institute for Nuclear Physics, and CNR—National Research Council, Italy.

\appendix

\section{Jordan-Wigner diagonalization of the Hamiltonian}
\label{App:JW_diagonalization}

 Quantum XY model described by the Hamiltonian in Eq.(\ref{Eq.Hamiltonian}) can be solved analytically by mapping spins into free fermions under the following Jordan-Wigner transformation \cite{barouch_pra_1970_1, barouch_pra_1970_2, lieb1961, glen2020}:
\begin{align}
    \sigma^x_n &=  \left( c_n + c_n^\dagger \right)
 \prod_{m<n}(1-2 c^\dagger_m c_m) \nonumber \\
    \sigma^y_n &=i\left( c_n - c_n^\dagger \right)
 \prod_{m<n}(1-2 c^\dagger_m c_m) \nonumber \\\text{and}\quad
\sigma^z_n&=1-2 c^\dagger_n  c_n,
 \label{eq:Jordan_wigner}
\end{align}
where \(c_m^\dag\)(\(c_m\)) is the creation (annihilation) operator of spinless fermions, and they follow the fermionic commutator algebra. 

We now apply the Fourier transformation of the fermionic operator $c_j$ and $c^{\dagger}_j$:
\begin{eqnarray}
    c_j = \frac{1}{\sqrt{N}}\sum_{p=-N/2}^{N/2} \exp[-i\frac{2\pi jp}{N}]c_p,\\
    c_j^{\dagger} = \frac{1}{\sqrt{N}}\sum_{p=-N/2}^{N/2} \exp[i\frac{2\pi jp}{N}]c_p^{\dagger},
\end{eqnarray}
The Hamiltonian in Eq.(\ref{eq:JW_hamil}) can then be block-diagonalized in the following way:
\begin{eqnarray}
    H &=& \sum_{p>0}^{N/2} (\sum_r J_r \cos(\phi_pr)+h)(c^{\dagger}_{-p}c_{-p}+c^{\dagger}_pc_p)\nonumber \\&-& i\gamma (\sum_r J_r \sin(\phi_pr))(c_pc_{-p}+c^{\dagger}_pc^{\dagger}_{-p}) - h \nonumber\\
    &=& \sum_{p>0}^{N/2} H_p
\end{eqnarray}
In the basis $\{\ket{0},c^{\dagger}_pc^{\dagger}_{-p}\ket{0},c^{\dagger}_p\ket{0},c^{\dagger}_{-p}\ket{0}\}$, we write down the matrix form of $H_p$:
\begin{widetext}
    \begin{eqnarray}
        H_p = 
        \begin{pmatrix}
            -h & i\gamma \sum_r J_r \sin(\phi_pr) & 0 & 0\\
            -i\gamma \sum_r J_r \sin(\phi_pr) & 2\sum_r J_r \cos(\phi_pr)+h & 0 & 0\\
            0 & 0 & \sum_r J_r \cos(\phi_pr) & 0\\
            0 & 0 & 0 & \sum_r J_r \cos(\phi_pr)
        \end{pmatrix}
    \end{eqnarray}
\end{widetext}

In a similar fashion, we can obtain the matrix forms of the blocks $\tau^{ij}_p$ of the two-qubit pauli operators $\tau^{ij}$ and the magnetization $\sigma^z$ written in momentum space $\sigma^z_p$:
\begin{widetext}
    \begin{eqnarray}
        \tau^{xx}_p = 
        \begin{pmatrix}
            0 & i \sin(\phi_p) & 0 & 0\\
            -i\sin(\phi_p) & 2\cos(\phi_p) & 0 & 0\\
            0 & 0 & \cos(\phi_p) & 0\\
            0 & 0 & 0 & \cos(\phi_p)
        \end{pmatrix};\nonumber
        \tau^{xy}_p = 
         \begin{pmatrix}
            0 & - \sin(\phi_p) & 0 & 0\\
            -\sin(\phi_p) & 0 & 0 & 0\\
            0 & 0 & \sin(\phi_p) & 0\\
            0 & 0 & 0 & -\sin(\phi_p)
        \end{pmatrix};\nonumber\\
        \tau^{yx}_p = 
        \begin{pmatrix}
            0 & - \sin(\phi_p) & 0 & 0\\
            -\sin(\phi_p) & 0 & 0 & 0\\
            0 & 0 & -\sin(\phi_p) & 0\\
            0 & 0 & 0 & \sin(\phi_p)
        \end{pmatrix};\nonumber
        \tau^{yy}_p = 
        \begin{pmatrix}
            0 & -i \sin(\phi_p) & 0 & 0\\
            i\sin(\phi_p) & 2\cos(\phi_p) & 0 & 0\\
            0 & 0 & \cos(\phi_p) & 0\\
            0 & 0 & 0 & \cos(\phi_p)
        \end{pmatrix};\nonumber\
    \end{eqnarray}
    \begin{eqnarray}
        \sigma^{z}_p = 
        \begin{pmatrix}
            -1 & 0 & 0 & 0\\
            0 & 1 & 0 & 0\\
            0 & 0 & 0 & 0\\
            0 & 0 & 0 & 0
        \end{pmatrix}\nonumber;
    \end{eqnarray}
\end{widetext}

\bibliography{ref}

\begin{thebibliography}{82}
\bibitem [{log()}]{log_neg}%
  \BibitemOpen
  \bibinfo {note} {The entanglement between subsystems \(i\) and \(j\) is
  quantified through the logarithmic negativity, defined for the bipartite
  density matrix \(\rho_{ij}\) as \begin{equation*} E(\rho_{ij}) = \log_2
  [\,2\mathcal{N}(\rho_{ij}) + 1\,], \end{equation*} where
  \(\mathcal{N}(\rho_{ij})\) denotes the sum of the absolute values of the
  negative eigenvalues of the partially transposed density matrix,
  \(\rho_{ij}^{T_i}\).}\BibitemShut {Stop}%
\end{thebibliography}

\end{document}